
\overfullrule=0pt
\font\twelvebf=cmbx12
\nopagenumbers
\line{\hfil CU-TP-539}
\line{\hfil FERMILAB-Pub-91/312-A\&T}
\vglue .5in
\centerline {\twelvebf  Black Holes in Magnetic Monopoles}
\vskip .3in
\centerline{\it Kimyeong Lee,$^*$  V.P.Nair}
\vskip .1in
\centerline { Physics Department, Columbia University}
\centerline {New York, New York 10027}
\vskip .2in
\centerline{\it and}
\vskip .2in
\centerline{\it Erick J. Weinberg}
\vskip .1in
\centerline{Physics Department, Columbia University}
\centerline{New York, New York 10027}
\centerline{and}
\centerline{Theory Group and NASA/Fermilab Astrophysics Center}
\centerline{Fermi National Accelerator Laboratory}
\centerline{P.O.Box 500, Batavia, Illinois 60510.}
\vskip .4in
\baselineskip=14pt
\overfullrule=0pt
\centerline {\bf Abstract}
\vskip .1in
We study magnetically charged classical solutions of a
spontaneously broken gauge theory interacting with gravity.  We
show that nonsingular monopole solutions exist only if the Higgs
field vacuum expectation value $v$ is less than or equal to a
critical value $v_{cr}$, which is of the order of the Planck
mass. In the limiting case, the monopole becomes a black hole,
with the region outside the horizon described by the
critical Reissner-Nordstrom solution.  For $v < v_{cr}$, we find
additional solutions which are singular at $r=0$, but which have
this singularity hidden within a horizon.  These have nontrivial
matter fields outside the horizon, and may be interpreted as
small black holes lying within a magnetic monopole.   The nature
of these solutions as a function of $v$ and of the total mass $M$
and their relation to the Reissner-Nordstrom solutions is
discussed.
\noindent\footnote{}{This work was supported in part by the US Department of
Energy (VPN and EJW), by NASA (EJW) under grant NAGW-2381 and by an NSF
Presidential Young Investigator Award (KL).}
\noindent\footnote{}{* Alfred P. Sloan Fellow.}
\vfil
\eject
\pageno 1
\input phyzzx
\overfullrule=0pt
\def\Int{\int_{r_H}^\infty}
\def\M{{\cal M}}
\def\high{\vphantom{\Biggl(}\displaystyle}
\catcode`@=11
\def\@versim#1#2{\lower.7\p@\vbox{\baselineskip\z@skip\lineskip-.5\p@
    \ialign{$\m@th#1\hfil##\hfil$\crcr#2\crcr\sim\crcr}}}
\def\simge{\mathrel{\mathpalette\@versim>}} %
\def\simle{\mathrel{\mathpalette\@versim<}} %
\catcode`@=12 

\def\pr#1#2#3#4{Phys. Rev. D{\bf #1}, #2 (19#3#4)}

\chapter { Introduction}

       Some spontaneously broken gauge theories contain magnetic
monopoles which have the remarkable property that, despite being
particles in a quantum theory, they are described by a
classical field configuration.   This is possible because in the
limit of weak gauge coupling $e$ their Compton wavelength $\sim
e/v$ is much less than the radius $\sim 1/(ev)$ of the classical
monopole solution.   Another curious property emerges as the
Higgs vacuum expectation value $v$ approaches the Planck mass
$M_P$.   The Schwarzschild radius $2MG \sim v/(eM_P^2)$ becomes
comparable to the monopole radius, suggesting that for $v \simge
M_P$ the monopole should be a black hole\rlap.\ref{J.A.Frieman and
C.T.Hill, SLAC preprint SLAC-PUB-4283 (1987).}
(This result can be
evaded in theories containing dilatons\rlap.\ref{J.A.Harvey and J.Liu,
University of Chicago preprint EFI-91-27 (1991).})
If $e \ll 1$, this
occurs in a regime where the energy density is much less than
$M_P$, justifying the neglect of quantum gravity effects.  By
studying the classical solution, then, one can gain
insight into how the particle passes over into a black hole.

        In this paper we undertake such an investigation.  We
consider an $SU(2)$ gauge theory in which a triplet Higgs field
$\phi$ breaks the symmetry down to $U(1)$;
\footnote{*}{ There has been some study
\ref{ see for example, D.V.Galt'sov and A.A.Ershov,
Phys.Lett. {\bf A138}, 160 (1989); M.S.Volkov and D.V.Galt'sov,
JETP Lett. {\bf 50}, 346 (1989); R.Bartnik and J.Mckinnon, Phys.
Rev.Lett. {\bf 61}, 141 (1988).} of black hole like
solutions in the theory without
Higgs fields.}
this theory gives rise
to 't Hooft-Polyakov monopoles with magnetic charge $Q_M = 1/e$.
We find that when $v = \langle \phi \rangle$ is sufficiently
large, the only magnetically charged solutions are the
Reissner-Nordstrom black holes.  These are essentially Abelian,
in that the only nontrivial matter field is the Coulomb magnetic
field lying in the unbroken $U(1)$ subgroup.  They have a
singularity at $r=0$ whose strength is determined by the mass
$M$. In order that this singularity be hidden within a horizon,
$M$ must be greater than $M_{crit} = \sqrt{4\pi Q_M^2} M_P$.  For
smaller values of
$v$, however, we find that a rather different type of black
hole solutions is also possible.  For these the horizon lies within the
core of the monopole, so that the non-Abelian structure is quite
evident in the region outside the horizon.   In a sense, these
solutions  can be viewed as black holes lying inside monopoles. The
mass of these objects can take any value down to the mass of the
nonsingular monopole.  We find that there is also an upper limit on
their mass. In some cases this limit is greater than the critical
Reissner-Nordstrom mass, so that there are two different black hole
solutions with the same values for the mass and magnetic charge.

     Classical solutions can also play another role in relation
to semiclassical gravity.  Black holes can reduce their mass by
Hawking radiation. By this mechanism initially macroscopic black
holes can shrink to the microscopic size characteristic of the
classical solutions. (In the weak coupling limit this scale is
much greater than the Planck length, so gravity can still be
treated semiclassically.)  Once this happens, the classical
solutions provide possible pathways for the further evolution of
the black hole by the Hawking process.  An understanding of the
nature of these solutions as a function of coupling constants and
other parameters can thus lead to further insight into the late
stages of magnetically charged black holes.

      In Sec.~2 we review the essential features of the theory
and obtain the field equations which must be obeyed by static
spherically symmetric solutions.  Much of this reproduces,
although with a somewhat different notation, the results of van
Nieuwenhuizen, Wilkinson, and Perry\Ref\vanN{P. van Nieuwenhuizen, D.
Wilkinson, and M.J. Perry, \pr{13}{778}76. }.  In Sec.~3 we study
nonsingular monopole solutions and their behavior as $v$
approaches $M_P$.  In Sec.~4 we extend our considerations to
include solutions with singularities inside the horizon which are
essentially black holes inside the monopole.
Section~5 contains some concluding remarks.  There are two
appendices.  In the first, we reconcile the absence of
nonsingular solutions for large $v$ with the existence of a positive
definite functional whose minima are solutions of the field equations.
In the second, we derive a number of inequalities which restrict the
properties of  the various types of black hole solutions.

\chapter{ General Formalism}

     The theory is governed by the action
$$ S = \int d^4x \sqrt{-g} \left[ -{1\over 16\pi G} R
   + {\cal L}_{matter} \right]
       \eqno\eq$$
where
$$ {\cal L}_{matter} =  -{1\over 4} (F^a_{\mu\nu})^2
          + {1\over 2} (D_\mu \phi^a)^2
        - {\lambda \over 2} ((\phi^a)^2 -v^2)^2
      \eqno\eq$$
with Latin indices $a,b, \dots$ referring to the internal $SU(2)$ indices,
$$  (D_\mu \phi)^a = \partial_\mu \phi^a - e \epsilon_{abc} A^b_\mu \phi^c
     \eqno\eq$$
and
$$  F^a_{\mu\nu} = \partial_\mu A^a_\nu - \partial_\nu A^a_\mu
         -  e \epsilon_{abc} A^b_\mu A^c_\nu
     \eqno\eq$$
A constant term has been included in the scalar field potential
so that the energy vanishes in the symmetry-breaking vacuum.  The
elementary excitations about this vacuum are a massless photon,
two charged vectors with mass $m_V = ev$, and a neutral massive
Higgs scalar with mass $m_H = 2 \sqrt{\lambda}v$.

    In this paper we consider only static spherically symmetric solutions.
For these, the metric may be written in the form
$$ ds^2 =  B(r)dt^2 - A(r)dr^2
    - r^2 (d\theta^2 + \sin^2 \theta d\phi^2)
     \eqn\metric$$
The normalization of $t$ is fixed by requiring that
$B(\infty)=1$, while the requirement that space be asymptotically
flat imposes the condition $A(\infty)=1$.  For later convenience
we define $\M(r)$ by
$$ A(r) = \left[ 1 -{2G\M(r)\over r}\right]^{-1}
     \eqno\eq $$

    For the matter fields we adopt the standard spherically symmetric ansatz
corresponding to magnetic charge $Q_M = 1/e$.
In flat space this ansatz is usually written in terms of
Cartesian coordinates as
$$ \phi^a = v {\hat r}^a h(r)
     \eqn\phiansatz$$
$$ A^a_i = \epsilon_{iak} {\hat r}^k\, {1 -u(r) \over er}
     \eqn\Aansatz$$
$$ A_0 = 0
     \eqn\Anaught$$
The extension to curved space\refmark{\vanN} is most easily done by first
transforming to spherical coordinates.
\footnote{*}{ An invariant way
to write the ansatz for the vector potential is
$$
A^{a}_{\mu}dx^{\mu}=f~ L^{a\mu}g_{\mu\nu}dx^{\nu}
$$
where $L^{a\mu}{{\partial}\over{\partial x^{\mu}}}$ are the three
Killing vectors corresponding to the rotational symmetry and $f$ is a
function invariant under the action of these Killing vectors. For the choice
of spherical coordinates in Eq.~\metric, this is the same as
Eqs.~\Aansatz\ and \Anaught.} Once this has been done, the matter part
of the action can be written as
$$ S_{matter} = -4\pi\int dt\, dr\, r^2\sqrt{AB}
     \left[ {K(u,h) \over A} + U(u,h) \right]
    \eqno\eq$$
where
$$ K = {{u^\prime}^2 \over e^2r^2} + {1\over 2}v^2 {h^\prime}^2
     \eqno\eq$$
and
$$ U = {(u^2-1)^2\over 2e^2r^4} + {u^2h^2v^2\over r^2}
    + {\lambda\over 2} v^4(h^2-1)^2
     \eqno\eq$$
(Primes denote differentiation with respect to $r$.)

$U(u,h)$ may be viewed as a position-dependent field
potential.  For later reference, we enumerate here its stationary
points:

     a) $u = \pm 1$, $ h =0  $: This is a local minimum of $U$
if $r<  1/(\sqrt{\lambda}v)$, and is a saddle point otherwise.

      b)  $u = 0$, $ h =0  $: This is always a local maximum of $U$.

    c) $ u = \hat u(r) $, $ h = \hat h(r)$, where
$$ \hat u(r) = \sqrt{ \lambda (1- e^2r^2v^2) \over \lambda -e^2}
    \eqn\uhat$$
$$ \hat h(r) = \sqrt{\lambda r^2v^2 -1 \over (\lambda -e^2) r^2v^2}
    \eqn\hhat$$
These are both real only when $rv$ lies between $1/|e|$ and
$1/\sqrt{\lambda}$.  Within this range of $r$, this point is the
global minimum of $U$ if $\lambda > e^2$, and a saddle point
otherwise.  When $\lambda = e^2$, $\hat u$ and $\hat h$ are
undefined, and this stationary point is replaced by a degenerate
set of minima, with $h^2 + u^2 =1$, which exist only when $|e|vr
=1$.

    d) $ u = 0$, $ h =\pm1 $: This is a local minimum of $U$ if
$r> 1/(ev)$, but only a saddle point otherwise.

     The gravitational field equations reduce to two independent
equations, which may be written as
$$ {(AB)^\prime \over AB} =16\pi G r K
     \eqn\ABeq$$
and
$$ \M^\prime = 4\pi r^2 \left( {K\over A} + U \right)=4\pi
r^{2}(K+U)~-~8\pi GrK \M
     \eqn\Meq$$
The equations for the matter fields are
$$ \eqalign{ {1\over \sqrt{AB}} \left( \sqrt{AB}
         u^\prime\over A\right)^\prime
    &= e^2{r^2 \over 2} {\partial U\over \partial u} \cr
    &= {u(u^2-1)\over r^2} + e^2uh^2v^2 }
    \eqn\ueq $$
and
$$ \eqalign{ {1\over r^2\sqrt{AB}} \left( r^2 \sqrt{AB}
         h^\prime\over A\right)^\prime
    &= {1\over v^2} {\partial U\over \partial h} \cr
    &= {2hu^2\over r^2} + 2\lambda h (h^2 -1)v^2 }
    \eqn\heq $$
By rescaling of distances in these equations, it can be shown
that that $u$, $h$ and $e\M/v$ are given by functions of $evr$
whose forms depend only on $\lambda/e^2$ and $Gv^2= (v/M_P)^2$.

   By integration of \Meq\  we see that the mass
can be written as
$$
M\equiv \M (\infty) ~=~
4\pi\int_{0}^{\infty}dr\,r^2 e^{-P(r)}\bigl( K+U\bigr)
   + e^{-P(0)} \M(0)
     \eqn\Eeq
$$
where
$$
P(r)= \int^{\infty}_{r}dr~8\pi GrK \eqno\eq
$$
If $\M(0) \ge 0$, the mass defined by \Eeq\ satisfies the
inequality\refmark{\vanN}
$$
M\geq  e^{-P(0)}
    \left[4\pi \int_{0}^{\infty}dr\,r^2 (K+U) + \M(0) \right]
    \geq {4\pi v\over e}~e^{-P(0)}
\eqno\eq
$$
which follows from the positivity of $K$ and the Bogomol'nyi
bound\rlap.\ref{E.Bogomol'nyi, Sov.J.Nucl.Phys., {\bf 24}, 449 (1975);
S.Coleman, S.Parke, A.Neveu, and C. Sommerfield, \pr{15}{544}77.}
For a nonsingular solution, $\M(0)=0$. A solution $u,h$
is then a minimum of $M$, viewed as a functional of $u$
and $h$, and so we have
$$
M(u,h)\leq M(u_{0}, h_{0})\leq
     4\pi\int_{0}^{\infty}dr\,r^2 (K+U)\vert_{u_{0},h_{0}}
 ~\leq M_{flat} \eqno\eq
$$
where $u_{0}$ and $h_{0}$ are the flat-space solutions and
$M_{flat}$ is the flat-space monopole mass. This inequality is of
course in accord with our intuition that gravity tends to reduce
the mass.

     Because the function space is noncompact, there is no guarantee
that there will actually be a configuration which minimizes $M$.
Instead, there could be an infinite sequence of configurations of
decreasing energy which does not converge on a limiting configuration.
Indeed, for $Gv^2$ sufficiently large we find that there are no
nonsingular solutions.  In Appendix~A we display a sequence of
configurations for which $M$ approaches, but does not reach, its lower
bound  for this case.

    Eq.~\ABeq\ can be used to eliminate $B(r)$ from the remaining
field equations, leaving one first-order and two second-order
equations to be integrated. A solution of these is determined by
five boundary conditions.  Two are provided by the
asymptotic conditions $u(\infty)=0$ and $h(\infty)=1$.  The
remaining three can be obtained by requiring that the solution be
nonsingular at the origin, which implies that
$u(0)=1$, $h(0)=0$, and $\M(0)=0$.

     Matters become more complicated if horizons are present, i.e.,
if $1/A(r)$ has zeros.  At a horizon $r_H$, the vanishing of
$1/A$ gives
$$ \M(r_H) = {r_H\over 2G}
    \eqno\eq $$
This, together with Eq.~\Meq, implies that
$$ \left({1\over A}\right)^\prime = {1\over r} - 8\pi G r U(u,h),
    \qquad r=r_H
    \eqn\ainvprime $$
Substitution of this into Eqs.~\ueq\ and \heq\ gives two
conditions
$$ u^\prime   \left[ {1\over r} - 8\pi G U(u,h) \right]
       ={e^2r^2 \over 2} {\partial U\over \partial u} ,
    \qquad r=r_H
     \eqn\uhoriz $$
and
$$ h^\prime   \left[ {1\over r} - 8\pi G U(u,h) \right]
      = {1\over v^2} {\partial U\over \partial h} ,
    \qquad r=r_H
    \eqn\hhoriz $$
which must hold if the solution is to be nonsingular.

     Since these additional conditions overdetermine the
solution, we do not expect there to be any nonsingular solutions
containing horizons, except perhaps for special values of
$\lambda$ and $v$.   If singularities are allowed at $r=0$,
then it should be possible to obtain otherwise nonsingular
solutions with a single horizon, but not (for generic values of
$\lambda$ and $v$) with two or more.

     Finally, note that a solution\ref{F.A.Bais and R.J.Russell,
\pr{11}{2692}75; Y.M.Cho and P.G.O.Freund, \pr{12}{1588}75.} of the
field equations is obtained by taking constant fields $u(r)=0$ and
$h(r)=1$, with  $$ \M(r) = M - {2\pi \over e^2r}
      \eqno\eq$$
where $M$ is arbitrary.
This yields the Reissner-Nordstrom metric
$$ B(r) = A(r)^{-1} = 1 -{2MG\over r}
        +  {4\pi G\over e^2r^2}
     \eqno\eq$$
which has a singularity at $r=0$.  There are horizons at
$$  r_\pm = MG \pm \sqrt{M^2G^2 - {4\pi G\over e^2}}
     \eqno\eq$$
provided that $M$ is greater than the critical value
$M_{crit}= \sqrt{4\pi/(Ge^2)}$.

\chapter{ Nonsingular Monopoles}

      For values of $v$ much smaller than the Planck mass,
gravitational effects on the monopole are small and the
nonsingular monopole solutions should be similar to their
flat-space counterparts.  On the other hand, as we have argued
above, if $v \gg M_P$, the Schwarzschild radius would be greater
than the size of the monopole, so the monopole must be a black
hole.  In this section we examine the transition between these two
regimes.

     Near the origin, a nonsingular solution must behave as
$$    u(r) = 1 - C_ur^2 + \cdots
      \eqn\smallru $$
$$    h(r) = C_hr + \cdots
      \eqn\smallrh $$
$$   \M(r) = {4\pi\over 3} \left( {6C_u^2 \over e^2}
     + {3\over 2}C_h^2v^2 + {\lambda\over 2}v^4 \right) r^3
      \eqno\eq $$
where $C_u$ and $C_h$ are constants which must be chosen so that
$u$ and $h$ approach the correct values as $r \rightarrow\infty$.
In the absence of gravity, the possibility of making such a
choice is ensured by the existence of a positive definite energy
functional whose minimum is a solution of the static field
equations.  This argument can be extended to the case of weak
gravity\refmark{\vanN} (see Eq. \Eeq), although, as discussed in
Sec.~2 and Appendix A, it fails when $v/M_P$ becomes too great.

     Just as in flat space, the matter fields $u$ and $h$ remain
nontrivial inside the monopole ``core'' of radius $\sim 1/(ev) $ and
then approach their asymptotic values exponentially fast:
$$    u(r) = O(e^{-m_Vr})
      \eqno\eq $$
$$    h(r) = 1- O(e^{-m_Hr})
      \eqno\eq $$
It then follows from Eq.~\Meq\ that
$$   \M(r)  = M -{2\pi \over e^2r} + O(e^{-m_Vr}, e^{-m_Hr})
       \eqn\Masym $$
In flat space, the monopole mass $M_{mon} \equiv M_{flat}= (4\pi v/e)
f(\lambda/e^2)$, where $f$ ranges from 1 to 1.787 as $\lambda$
ranges from 0 to $\infty$\rlap.\ref{T. Kirkman and C. Zachos,
\pr{24}{999}81.}\  As mentioned above, gravitational effects cause
$M$ to be somewhat smaller; our numerical results indicate that the
monopole mass can be reduced to about two-thirds of its flat-space
value.

      We now turn to the discussion of how a horizon develops
as the mass increases. From the large and small $r$ behavior of
$\M$, it is evident that $1/A$ will have a minimum,
corresponding to a maximum of $\M/r$, at some intermediate value
of $r$.  The asymptotic form Eq.~\Masym\ suggests that this occurs
at a value $\bar r\sim 4\pi /(e^2M) \sim 1/(ev)$, with $1/A(\bar
r) \approx 1 - O(Gv^2)$.  As $v$ increases, this minimum
should become deeper, until eventually a critical value $v_{cr}$
is reached for which $1/A(\bar r) =0$ and a horizon appears.
One would expect this horizon to persist if $v$ were increased
further, but, as was argued in the previous section, it will not
in general be possible for a solution with proper asymptotic
behavior to be well-behaved at both the horizon and at $r=0$.  We
therefore expect that only singular solutions exist when $v >
v_{cr}$. More specifically, our results for the critical case
suggest that in the supercritical case there are only
Reissner-Nordstrom solutions with $u(r)=0$ and $h(r)=1$; in the
next section we will prove this to be the case if $v$ is
sufficiently great.

    Let us examine the critical case $v=v_{cr}$ in more detail.  To
begin, note that Eqs.~\uhoriz\ and \hhoriz, together with the fact
that $1/A$ is stationary at the horizon, imply that $u(r_H) $
and  $h(r_H) $ must correspond to one of the
stationary points of $U(u,h)$, which were enumerated in Sec.~2
The first two, $u = \pm 1$, $h =0  $ and $u = 0$, $ h
=0$, are easily ruled out.  For the former, one can show that if
$1/A$ and $(1/A)^\prime$ both vanish, then $(1/A)^{\prime\prime}$
must be negative, in contradiction with the assumption that $1/A$
is at a minimum. In the latter case, it is easy to show that
all solutions of Eq.~\ueq\ and \heq\ develop singularities
as $r \rightarrow r_H$ if $(1/A)^{\prime\prime} >0$.

     We have not been able to completely eliminate the third
case, $u = \hat u(r), h=\hat h(r)$.  There are however
several constraints which the parameters  must satisfy for a
solution to exist\rlap.\ref{P.Hajicek, Proc. Roy. Soc. {\bf A 386},
223 (1983) and J.Phys. A {\bf 16}, 1191 (1983).}  The condition $(1/
A)'=0$ implies $4\pi G(u^2 +h^2 )=1$ at the horizon.
With the values of $\hat u(r),~\hat h(r)$ from Eqs.~\uhat\ and \hhat,
this leads to a quadratic equation for $r_H^2$.  The requirements that
$vr_H$ lie between $1/e$ and $1/\sqrt{\lambda}$, so that $\hat u$ and
$\hat h$ are both real, and that $(1/A)^{\prime\prime} >0$ eliminate
one of the solutions of the quadratic equation and lead to the
conditions
$$ \eqalign{ & 1 + \sqrt{e^2\over \lambda}
    \, \ge\, 8\pi Gv^2   \,\ge\, 2, \qquad  \lambda < e^2 \cr
     &1 + \sqrt{e^2\over \lambda}
    \,\le\, 8\pi Gv^2  \, \le\, 2, \qquad  \lambda > e^2 }
    \eqno\eq$$
In addition to these requirements, the solution in the region
within the horizon must be such that $2G\M(r_H) = r_H$; we do
not know whether this can be done with $v$ in the range
specified above.  Furthermore, we have not addressed the
question of whether these solutions are stable; this seems
particularly doubtful for the case $\lambda < e^2$, where
$\hat h$, $\hat u$ is not a minimum of $U(h,u)$.

      Finally, we come to the case $u=0$, $h=1$. This corresponds
to a solution in which $u$ and $h$ have already
reached their asymptotic values at the horizon which, from
Eq.~\ainvprime,  must occur at
$$  r_H =  \sqrt{4 \pi G\over e^2}
   \eqno\eq$$
The entire monopole, except for its Coulomb magnetic field, lies
within the horizon.  The exterior solution is then of the
Reissner-Nordstrom form with the mass $M$
equal to the critical value for unit magnetic charge.  Since we
want it to be nonsingular, the interior solution cannot be simply
Reissner-Nordstrom.  Instead, it is similar in form to the solutions
for subcritical $v$ at small $r$, while near the horizon $u$ and
$1-h$ vanish as powers of $r_H-r$.   Two aspects of this solution
may seem puzzling.  First, it may seem unphysical for the entire
evolution of the matter fields to take place within a finite
range of $r$. However, this becomes more plausible when one notes
that the physical distance from the origin is
$$  l(r) = \int_0^r dr\, \sqrt{A(r)}
    \eqno\eq $$
Since $A$ diverges as $(r - r_H)^{-2}$ near the horizon of the
critical solution, $l(r_H)$ is in fact infinite.  In a sense,
rather than the monopole being compressed to fit within the
horizon, the horizon has been expanded outward to encompass the
monopole.   Second, the values for the fields and their derivatives
at any $r > r_H$ do not determine the solution everywhere, as
evidenced by the fact that this solution and the Reissner-Nordstrom
agree in the exterior region but differ in the interior.  This is
possible because the simultaneous vanishing of $1/A$ and
$(1/A)^\prime$ at the horizon prevents one from simply integrating
across the horizon and allows nonanalytic behavior at $r=r_H$.

      We have checked these arguments by numerically solving the
field equations.  Starting with the small distance expansions of
Eqs.~\smallru\ and \smallrh, we varied the constants $C_u$ and
$C_h$ until the proper asymptotic behavior was obtained.  In all
cases we found that as $v$ approached $v_{cr}$ the solution
tended toward one which was purely Reissner-Nordstrom in the
exterior region, rather than one for which the fields were given
by $\hat u$ and $\hat h$ at the horizon.

\FIG\plotfig{Plots of (a) $u(r)$, (b) $h(r)$, and (c) $1/A(r)$ for
$\lambda/e^2 = 1.0$ and $\mu= 8\pi Gv^2$ equal to 0.1 (solid line), 1.0
(dashed-dotted line), 2.0 (dotted line), and 2.35 (dashed line).}

\FIG\physdistfig{Plot of $u(r)$ as a function of $l(r)$, the physical
distance from the origin, for $\mu= 8\pi Gv^2$ equal to 0.1  (solid
line), 1.0 (dashed-dotted line), 2.0 (dotted line), and 2.35 (dashed
line).}

     A sample of these results is displayed in Fig.~\plotfig, where
we show $u$, $h$, and $1/A$ as functions of $r$ for $\lambda/e^2
= 1.0$ and $\mu= 8\pi Gv^2$ equal to 0.1, 1.0, 2.0, and 2.35.
The last of these values is as close as we were able to come to
the critical value $\mu_{cr}= 8\pi Gv^2_{cr}$.  As $\mu$ increases,
the monopole appears to be pulled inward.  The minimum of $1/A$
also moves inward, although less so.  A contrasting view is obtained
by plotting these fields as functions of the physical distance $l(r)$.
As an example, $u(r)$ is plotted in this fashion in Fig.~\physdistfig;
we see that the change in the physical size of the monopole is
actually rather small.

    We also studied the behavior of $\mu_{cr}$, finding it to be a
decreasing function of $\lambda/e^2$.  In particular, for $\lambda/e^2$
equal to 0.1, 1.0, and 10.0, $\mu_{cr}$ is 3.7, 2.4, and 1.6,
respectively.

\chapter{Black Holes in Monopoles}

        It was argued in the previous section that for $v >
v_{cr}$ all solutions will have singularities.
In this section we consider these supercritical
solutions as well as another class of singular solutions which may
be viewed as black holes embedded inside monopoles.  Let us suppose
that  $\M(0)$ is nonzero and positive, with $2G\M(0)$ much smaller
than the monopole radius, and that $v \ll M_P$, so that the
monopole would not  by itself become a black hole.  At
small $r$, the effects of the matter fields can be neglected and the
metric will be similar to that of a Schwarzschild black hole with
mass $\M(0)$.  At larger $r$, the gravitational effects will be
small and the matter fields will resemble those of a flat space
monopole.  One might object that having structure outside the
horizon would be forbidden by the no-hair theorems, and that the
monopole would collapse into a Reissner-Nordstrom black hole.
This is not so.
The behavior of the fields at the outer edges of the monopole core is
determined largely by the shape of the position-dependent field
potential $U$ at that radius.  The effects on the fields in this
region of a small black hole near the center of the monopole would be
small, much as the effects of a small black hole at the center of  a
large solid body (e.g., the Earth) would be neglible at the outer
regions of the body.  To understand how equilibrium is possible at the
horizon, note that at $r=r_H$ the covariant conservation of a diagonal
energy-momentum tensor reduces to the condition $\rho + p_r =0$, where
$T^\mu_\nu \equiv {\rm diag}( - \rho, p_r, p_\theta, p_\phi)$.  While
this cannot be achieved in normal fluids, which have positive
pressure, it is quite possible in a field theory.

      Let us now try to make these arguments more quantitative. We
begin by recalling the derivation of a no-hair theorem\ref{
J.D.Bekenstein, \pr5{1239}72.}
for a theory with a single scalar field $\phi$.   We restrict our
consideration to spherically symmetric configurations, so the matter
field equation can be written as
$$  \left( {r^2 \sqrt{AB} \phi^\prime \over A} \right)^\prime
   = r^2 \sqrt{AB} {dV \over d\phi}
    \eqno\eq$$
with the metric given by Eq.~\metric. Multiplying both sides by
$(\phi -\phi_0)$, where $\phi_0$ is a minimum of the $V(\phi)$,
and integrating from the horizon out to infinity gives
$$    \int^\infty_{r_H} dr\, r^2 \sqrt{AB}
      \left[ {1\over A}{\phi^\prime}^2 +
      (\phi-\phi_0){dV\over d\phi} \right]
   = \left. {r^2 \sqrt{AB} \phi^\prime (\phi-\phi_0) \over A}
      \right|^{r=\infty}_{r=r_H}
      \eqn\nohair$$
The right hand side vanishes, since $1/A(r_H)=0$, while energetic
arguments require that $\phi^\prime (\phi-\phi_0)$ fall faster
than $r^{-2}$ at large distances.  The first term in the integral
on the left hand side can never be negative, since $A(r) >0$
outside the horizon.  If $\phi_0$ is the only minimum of
$V(\phi)$, then the second term in the integrand is also
non-negative everywhere, and Eq.~\nohair\ can only be satisfied if
$\phi(r)=\phi_0$ for all $r>r_H$.  Thus a necessary condition for
the existence of a nontrivial field outside the horizon is that
$V(\phi)$ have more than one minimum.

     As we have seen, the monopole problem, when restricted to
spherically symmetric configurations, resembles a theory with two
scalar fields and a position-dependent field potential.  The
fields which minimize this potential are different at large and
small values of $r$.  If $2G\M(0)$ is much less than both
$1/(ev)$ and $1/(\sqrt{\lambda v}) $, we  would expect the fields at
the horizon to be at or near the short  distance minimum, $u=1$, $h=0$.
There would then be nontrivial  behavior in the region outside the
horizon as the fields evolved  to the asymptotic values
corresponding to the large distance  minimum.  On the other hand, if
the horizon is located at large  $r$, where $u=0$, $h=1$ is the only
minimum of the potential, the  no-hair theorem derived above
suggests that the fields must lie  at their asymptotic value
everywhere outside the horizon.

     This picture can be made more precise with the aid of certain
inequalities which $r_H$ and the values of the fields must obey, if
we make a few plausible assumptions.  We
assume that the fields vary monotonically outside the horizon, so
that $u'$ is everywhere negative and $h'$ is everywhere positive,
with $u$ and $h$ always taking values between zero and one.  At the
horizon, we have $(1/A)' \ge 0$, with equality holding only for
the critical solutions discussed in the previous section. From
Eqs.~\ainvprime-\hhoriz, we then have
$$
u^*(1-{u^*}^2 )\geq u^*{h^*}^2 e^2 v^2 r_{H}^2 \eqn\jeq
$$
$$
h^*{u^*}^2 \geq h^* (1-{h^*}^2 )\lambda v^2  r_{H}^2 \eqn\keq
$$
where $u^*\equiv u(r_H)$ and $h^*\equiv h(r_H)$.
Inequalities \jeq\ and \keq\ translate into the following
possibilities. Either $u=0,~h=1$ (corresponding to the exterior
solution being Reissner-Nordstrom) or
$$
\lambda v^2 r_{H}^2 (1-{h^*}^2 )\leq {u^*}^2 \leq 1-{h^*}^2 e^2 v^2
r_{H}^2 \eqn\uIeq
$$
$$  1 -{{u^*}^2\over \lambda r_H^2v^2} \le {h^*}^2
     \le {1-{u^*}^2 \over e^2r_H^2v^2}
     \eqno\eq$$
(There is one more possibility, viz. $h^*=0$. This can
only occur if $(1/A)'=0$, but we have already seen that for the
critical case $h^*=0$ leads to singularities at the horizon.)
Since $u^2$ and $h^2$ must lie between zero and one, these
inequalities require $$
\lambda v^2 r_{H}^2 \leq 1~~~~~~~~~~~~~~\lambda\leq e^2
$$
$$
e^2 v^2 r_{H}^2 \leq 1~~~~~~~~~~~~~~\lambda >e^2 \eqn\IIeq
$$
Since $r_{H}$ is given by $2G\M (r_{H})$, it clearly increases as
either  $\M(0)$ or $v$ is increased.   Thus if  we increase either
of these quantities, we will eventually
reach a point when these inequalities can no longer be satisfied.
When this happens, the only admissible solution to the inequalities
\jeq\ and \keq\ is $u^*=0$, $h^*=1$.

      One could summarize these results by drawing a ``phase
diagram'' of the solutions as a function of $\M(0)$ and $v$.   The
nonsingular solutions considered in the previous section would lie
along the $\M (0)=0$ axis, with $v<v_{cr}$.  Above this axis, and
to the left of a critical line, would be the black hole solutions
we have just described.  To the right of this line there would be
no solutions.  The Reissner-Nordstrom solutions would not appear on
the phase diagram, because for these $\M(0)= - \infty$.
diagram, because for these $\M(0)$ is infinite.

      Obtaining the precise boundaries in this phase diagram would
require that we return to the field equations \Meq-\heq\
and look for numerical solutions for various values of $\M(0)$
and $v$.   However, considerable insight can be gained by the
analysis of a somewhat simplified model of a monopole.
In this model the flat space
monopole is composed of a core of radius $R$ with uniform
energy density, with only the Coulomb magnetic field
extending outside the core.  The energy density is then
$$ \rho = \cases{ \high \rho_0,& $ r <R$\cr
          \high{1\over 2e^2r^4}, &$ r >R$ }
     \eqn\rhomodel$$
Integrating this to obtain the monopole mass $M_{mon}$, and then
minimizing with respect to $R$, gives
$$ R = \left( {1\over 2 e^2\rho_0} \right)^{1/4}
   \eqno\eq$$
and
$$  M_{mon} = {8\pi \over 3 e^2 R}
   \eqn\Mmoneq$$
with one fourth of the monopole mass lying within the core.
These results are in qualitative agreement with the exact results
if $R \sim 1/ev$, $\rho_0 \sim e^2 v^4$, and $M_{mon} \sim v/e$.

      We now use this model to calculate $\M(r)$ and then
use the result to determine the positions of the horizons.
Specifically, in the presence of gravity we define  $\rho
\equiv K/A + U =  \M^\prime/(4\pi r^2)$, and continue to model
it by Eq.~\rhomodel, with $R$ and $M_{mon}$ as given above.
This gives
$$  {\M(r) \over r} =
   \cases{ \high{\M(0)\over r} + {M_{mon} r^2\over 4 R^3} , &$ r< R$ \cr
     \high{\M(0) +M_{mon} \over r} - {3M_{mon} R \over 4 r^2} ,&$ r> R$}
     \eqno\eq$$

    The behavior of this function depends on the relative
magnitudes of $\M(0)$ and $M_{mon}$.
If $\M(0) < M_{mon}/2$, $\M(r)/r$ diverges at $r=0$, falls to a
minimum at $r_1 = (2\M(0)/M_{mon})^{1/3}R$, and then rises to a
maximum at
$$ r_2 = {3RM_{mon}\over 2 \left[\M(0) + M_{mon}\right]}
    \eqno\eq$$
with
$$  {\M(r_2) \over r_2} = {e^2 \over 8\pi}
    \left[ \M(0) + M_{mon} \right]^2
     \eqno\eq$$
It then decreases monotonically to zero as $r\rightarrow
\infty$.  The horizons occur at the values of $r$ such that $2G\M/r =
1$.  One such lies at a position $r_H < r_1$ such that
$$  {M_{mon}r_H^3\over 4R^3} + \M(0) = {r_H\over 2G}
    \eqn\firsthoriz$$
With small $v$ (and hence small $M_{mon}$), the peak at $r_2$ is
less than $1/(2G)$, and this is the only horizon.  As $v$ is
increased, with $\M(0)$ held fixed, the peak at $r_2$ rises, reaching
$1/(2G)$ when
$$ M_{mon} + \M(0) = \sqrt{ {4\pi \over e^2 G}}
     \eqno\eq $$
or, equivalently,
$$ M = M_{crit}
    \eqno\eq$$
where $M =\M(\infty)$ and we have introduced $M_{crit}$, the critical
Reissner-Nordstrom mass for unit magnetic charge.  This behavior is
quite analogous to that we saw for the $\M(0)=0$ case.  Just as in
that case, non-Reissner-Nordstrom solutions are not expected to exist
beyond this critical point.

    If instead $\M(0) > M_{mon}/2$, $\M(r)/r$ decreases
monotonically.  Taken at face value, our
formulas would always imply the existence of a horizon.
However, our discussion of the no-hair
theorem suggests that for a non-trivial solution to exist
the horizon must lie within the monopole core, in which case it must
satisfy Eq.~\firsthoriz.  Requiring that this equation have a solution
with $r_H < R$, and using Eq.~\Mmoneq, we obtain the condition
$$    M < {3\over 4} M_{mon}  + {M_{crit}^2 \over 3M_{mon}}
     \eqn\boundaryeq $$

\FIG\phasediag{The phase diagram of solutions for the simplified
monopole model discussed in the text.  ``R-N'' refers to a
Reissner-Nordstrom solution with a horizon, while ``Mon'' refers to
the solutions with a black hole inside a nontrivial monopole
configuration.}

     We can now construct the phase diagram of solutions.  This is
shown in Fig.~\phasediag, where we have labeled the axes by $M_{mon}$
(which is proportional to $v$) and $M$; we have chosen the latter
variable rather than $\M(0)$ in order to be able to include the
Reissner-Nordstrom solutions.  The line $OA$ is given by $M =
M_{mon}$, while the line $BC$ is determined by Eq.~\boundaryeq.  The
nonsingular monopole solutions lie along the line $OA$, with the
critical solution at point $A$.  In the region above and to the left
of this line, but below the line $ABC$, are the solutions with black
holes inside nontrivial monopole configurations.  Reissner-Nordstrom
solutions occur everywhere above the line $M= M_{crit}$.  These two
regions overlap to the left of $BC$; in this portion of the diagram,
there are two distinct solutions with the same values for $M$ and
$M_{mon}$.  Finally, since we are excluding solutions with naked
singularities, there are no solutions in the region to the right of $OA$
with $M < M_{crit}$.

\chapter{ Discussion}

       We have seen that a variety of black hole solutions may be
associated with the magnetic monopoles of spontaneously broken
gauge theories.  The Reissner-Nordstrom solutions with Abelian
magnetic charge have long been known; these need only a trivial
modification to accomodate the Higgs field.  A notable feature of
these is that they require a nonzero minimum mass for any given
magnetic charge.   The new class of solutions we have found can
have any mass down to that of the monopole, while the mass within
the horizon can be arbitrarily small.  Nevertheless, the black
hole certainly carries unit topological charge, since the Higgs
field is topologically nontrivial on the horizon.   Whether or
not it contains unit magnetic charge is somewhat less clearcut,
since the horizon lies in a region where the asymptotic
symmetry-breaking vacuum has not yet been established and where
the definition of the electromagnetic field strength is ambiguous.

      It is interesting to consider the evolution of these solutions
as the system moves in the $M$-$M_{mon}$ plane.
Since $v$ is  a constant of nature (although one might perhaps
envision a time-dependent $v$ in a cosmological context), this motion
must be along vertical lines in the phase diagram of  Fig.~\phasediag.
Accretion of incident external particles would
increase $M$ and move the system upward.   Downward motion could
arise spontaneously through Hawking radiation.   In particular, a
pure Reissner-Nordstrom solution has a Hawking
temperature\ref{N.D.Birrell and P.C.W.Davies, {\it Quantum Fields in
Curved Space}, Cambridge University Press (1982).}
$$ T = {1\over 2\pi G} {\sqrt{M^2 -M_{crit}^2} \over
   \left( M + \sqrt{M^2 -M_{crit}^2}\right)^2 }
      \eqno\eq $$
As this black hole radiates it loses mass and increases its
temperature, thus accelerating the mass loss, until it reaches a
maximum temperature
$$ T_{max} = {2 \over 3\sqrt{3}(4\pi)^{3/2} }e M_P
     \eqno\eq $$
when $ M = (2/\sqrt{3})M_{crit}$.  From this point, $T$ rapidly falls,
reaching zero when $M=M_{crit}$.   In the usual analysis, the
critical solution is thus the stable asymptotic endpoint of the
Hawking process, unless the black hole has managed to discharge
its magnetic charge \ref{W.A.Hiscock, Phys.Rev.Lett. {\bf 50}, 1734
(1983).};
by choosing $e$ small enough
this can be suppressed. However, our results suggest that
if $v < v_{cr}$ this may not be the whole story.
For the solutions we found in section 4, corresponding to black holes
inside monopoles, the radius of the horizon can be easily shown,
using \boundaryeq, to be larger than the horizon radius for the
Reissner-Nordstrom solution of the same mass. Classically, since the
area of the horizon cannot decrease, this suggests that
the Reissner-Nordstrom solutions are unstable, possibly decaying to our
solutions.
This can indeed be shown by a perturbation analysis
around the Reissner-Nordstrom solutions.\ref{ K.Lee, V.P.Nair and E.J.
Weinberg, Columbia-Fermilab preprint CU-TP-540/FERMILAB-Pub-91/326-A\&T}
There is thus the possibility of a transition
from the pure Reissner-Nordstrom solution to one
in which the horizon lies within the monopole core.
Once this
transition has occured,
there is no longer any obstacle to the complete evaporation
of the horizon.
These possibilities await further exploration.
\vskip .2in
\centerline{\bf Acknowledgements}

     We thank Joshua Frieman and Hai Ren for helpful
conversations.

\Appendix{A}

     If $\M(0)=0$, any minimum of the functional
$$    M = 4\pi\int_{0}^{\infty}dr\,r^2 e^{-P(r)}\bigl( K+U\bigr)
   + e^{-P(0)} \M(0)
     \eqno\eq $$
gives a nonsingular solution of the field equations.  Here
$$ P(r)= \int^{\infty}_{r}dr~8\pi GrK
     \eqno\eq $$
while $K$ and $U$ are the gradient and potential terms given by
Eqs.~(2.11) and (2.12).  We have seen that for $v > v_{cr} \sim M_P$
there are no nonsingular solutions, and hence no configuration which
minimizes $M$. Since $M$ is bounded from below (see Eq.~(2.21)), this
implies that there must be a sequence of configurations of decreasing
energy which does not converge on a limiting nonsingular configuration.
In this appendix we will display such a sequence.

    $M$ differs from the flat space energy functional by containing
the factor of $e^{-P}$.  Because of this factor, a rapid variation of
the fields $u$ and $h$ about some value $r=R$ leads to a
suppression of the integrand in the region $r<R$.  This suggests that
we consider configurations of the form
$$ u(r) = \cases{ 0, &$r> R+ {\Delta\over 2}$ \cr
                   f_u(r), &$|r-R| < \Delta$ \cr
                   1, &$r< R - {\Delta\over 2}$\cr }
    \eqno\eq$$
$$ h(r) = \cases{ 1, &$r> R+ {\Delta\over 2}$ \cr
                   f_h(r), &$|r-R| < \Delta$ \cr
                   0, &$r< R - {\Delta\over 2}$ \cr}
    \eqno\eq$$
where $f_u(r)$ and $f_h(r)$ are smooth functions interpolating between
the small $r$ and large $r$ values of the fields, and the limit
$\Delta \rightarrow 0$ will eventually be taken.  The large distance
values $u=0$, $h=1$ are chosen to minimize the potential term $U$; as
will become evident shortly, the precise choice of the short distance
values has no effect on the final result.  For configurations of this
form, $K$ vanishes everywhere except in the transition region $|r-R| <
\Delta$, where it is proportional to $1/\Delta^2$.  It follows that
$P(r) = 0$ for $r>R+\Delta/2$ and is proportional to $1/\Delta$ for $r
< R -\Delta/2$.  Hence, the entire contribution to $M$ from the
interior region is suppressed by a factor of the form $e^{-{\rm
const.}/\Delta}$.  The contribution from the exterior region, which is
due entirely to $U=1/(2e^2r^4)$, is simply $2\pi/(e^2(R+\Delta/2))$.
In the transition region, the contribution from $U$ is clearly of
order $\Delta$, while that from $K$ can be estimated by writing $r^2 =
r(R + O(\Delta))$ and noting that the leading part of the integrand is
then a total derivative.  This gives
$$ M = {R \over 2G} + {2\pi \over e^2R} + \cdots
     \eqno\eq$$
where the terms represented by dots are suppressed either
exponentially or by powers of $\Delta$ as $\Delta \rightarrow 0$.
Minimizing with respect to $R$ gives $R= \sqrt{4\pi G/e^2} + \cdots$
and  $M = M_{crit} + \cdots$.  As $\Delta$ tends to 0, $M$ approaches
the critical Reissner-Nordstrom mass $M_{crit}$, but the limiting
configuration, with $\Delta = 0$, is singular at $r=R$ and thus is
not an acceptable solution of the field equations.

\Appendix {B}

      In this appendix we derive some inequalities which apply
to solutions with horizons which are not necessarily Reissner-Nordstrom
outside the horizon.  In particular, these apply to the solutions,
considered in Sec.~4, which described black holes inside
monopoles.  We assume that for all $r\ge r_H$ the matter fields
$u$ and $h$ are nonsingular and take values between 0 and 1, that
$u$ is monotonically decreasing, and that $h$ is monotonically
increasing.

      We first derive bounds on the mass outside the horizon.
The first step is to note that, after eliminating $B$ with the aid
of Eq.~\ABeq,  Eqs.~\ueq and \heq for the matter fields can be written as
$$ \eqalign{
\left( {u'\over A}\right)'
 &~=~ {e^2r^2\over2} {\partial U\over \partial u}
    ~-~{{8\pi GrKu'}\over A} \cr
&~=~ {{u(u^2 -1)}\over {r^2}}~+~ e^2uh^2v^2~-~{{8\pi GrKu'}\over A} }
\eqn\modueq
$$
$$ \eqalign{
\left({{r^2 h'}\over A}\right)'
&~=~ { r^2\over v^2} {\partial U\over \partial h}
       ~-~ {{8\pi Gr^3 Kh'}\over A} \cr
&~=~ 2hu^2 ~+~ {r^2\over v^2} {{\partial V}\over
{\partial h}}~-~ {{8\pi Gr^3 Kh'}\over A} }
\eqn\modheq
$$
(For brevity, we have written $V$ for the Higgs potential
${\lambda\over 2}v^{4} (h^{2}-1)^{2}$.)  Integrating the first of these,
and recalling that $u'(\infty)= 1/A(r_H) =0$, we get
$$
\Int~dr~\left[e^2v^2 uh^2 ~-~ {{8\pi GrKu'}\over A}\right]
{}~=~ \Int~dr~{u(1-u^2 )\over {r^2}}
\eqn\eqAppi
$$

     We now integrate Eq.~\Meq\ to obtain the expression
$$
M-\M (r_{H})=4\pi \Int~dr~r^2 \left({K\over A}+U\right)
\eqn\outsidemass
$$
for the mass outside the horizon. Integrating by parts the $u'^2$
and $h'^2$ terms in $K$ and using the field equations \modueq\ and
\modheq, we obtain
$$ \eqalign{
M-\M(r_{H}) &=4\pi \Int~dr~\left[{{1-u^4}\over{2e^2 r^2}}~+~
r^2 V~+~ r^2 {h\over 2}\left(-{{\partial V}\over {\partial h}}+
{{8\pi GrKh'}\over A}\right) \right. \cr
 & \qquad  \left.- u\left(uh^2v^2 - {{8\pi GrKu'}\over e^2 A}\right)
      \right] }
\eqn\eqAppii
$$
Since $u(r)\leq u(r_{H})\equiv u^*$ for $r\geq
r_{H}$, we have, using Eq.~\eqAppi,
$$ \eqalign{
\Int dr~ u\left( uh^2v^2 - {{8\pi GrKu'}\over e^2 A}\right)
  &\leq u^* \Int dr~ \left( uh^2v^2 - {{8\pi GrKu'}\over e^2 A}\right)\cr
 & \leq {u^*}^2 \Int dr~ {{(1-u^2)}\over e^2r^2} }
\eqno\eq
$$
Substitution of this into Eq.~\eqAppii leads to
$$ \eqalign{
M-\M(r_{H}) &\geq 4\pi \Int~dr~
    \left[ {1-u^4 - 2{u^*}^2(1-u^2) \over 2e^2r^2} \right] \cr
    &\ge 4\pi \Int~dr~  {1 -2{u^*}^2  \over 2e^2r^2} }
     \eqno\eq $$
and hence
$$M-\M(r_{H}) \geq { 2\pi \over e^2 r_H} (1 -2{u^*}^2)
     \eqn\lowerbound $$

       To get an upper bound on $M - \M(r_H)$, we start with the
identity
$$ \Int dr\, {d\over dr} (r^3U) = \Int dr\, \left[ r^2U
  + r{\partial \over \partial r}(r^2U)
  + r^3 u'{\partial U\over \partial u}
  + r^3 h'{\partial U\over \partial h} \right]
       \eqno\eq $$
The left hand side gives only a surface term at $r=r_H$. (The term at
$r=\infty$ vanishes.)  On the right hand side, the last two terms can
be rewritten with the aid of the field equations \modueq\ and \modheq.
After some algebra and an integration by parts this gives
$$ \eqalign{ \Int dr\, r^2U &=
   - r^3_H U(r_H) \cr
  -\Int dr\,&\left[ r{\partial \over \partial r}(r^2U)
     + {16\pi G r^4 K^2 \over A} + 2r^3K \left({1\over A}\right)'
     + {2r u' u'' \over e^2 A} + {rv^2 h'(r^2h')' \over A} \right] \cr
   &= - r^3_H U(r_H) \cr
  +\Int dr\,&\left[ {(1-u^2)^2\over e^2r^2} -2r^2V
     - {16\pi G r^4 K^2 \over A} - 2r^3K \left({1\over A}\right)'
     - {r^2 K\over A} + {{2u'}^2 \over e^2A} \right]}
        \eqno\eq $$
Inserting this into Eq.~\outsidemass\ and dropping positive terms, we
obtain $$  M- \M(r_H) \le 4\pi \left \{ - r^3_H U(r_H)
    +\Int dr\, \left[ {{2u'}^2 \over e^2A}
          + {(1-u^2)^2\over e^2r^2} \right] \right\}
          \eqno\eq$$
We now need a bound on the ${u'}^2$ term.  To obtain this we multiply
Eq.~\modueq\ by $u$ and integrate from the horizon to infinity to obtain
$$ \eqalign{
\Int dr\,  {u'^2 \over A}
     &= \Int dr\, \left[ {u^2 (1-u^2 )\over r^2} - e^2 v^2 h^2u^2 +
       {8\pi GrKuu' \over A} \right]\cr
     &\le \Int dr~ {{u^2 (1-u^2 )}\over r^2} }
    \eqno\eq $$
Hence,
$$  M- \M(r_H) \le 4\pi \left \{ - r^3_H U(r_H)
    +\Int dr\,{(1-u^4)\over e^2r^2} \right\}
          \eqno\eq$$
Dropping the $u^4$ in the integrand gives the inequality
$$
M-\M(r_H )\leq {2\pi \over {e^2 r_H}} (1+2u^{*2}-u^{*4})
      \eqn\upperbound $$
where $u^* \equiv u(r_H)$.
The inequalities \lowerbound\ and \upperbound\ can be combined as
$$
{2\pi \over {e^2 r_H}}(1-2{u^*}^2)
    \leq M-\M(r_H )\leq {2\pi \over {e^2 r_H}}
(1+2{u^*}^2 -{u^*}^4 )\eqn\bothbounds
$$

As discussed in text, when the mass is large enough so that the
inequalities (4.7) are no longer respected, we have Reissner-Nordstrom
solutions.
In this case $u^* =0$ and the inequalities \bothbounds\ simply say that
the mass outside the horizon is given by $2\pi/ {e^2 r_H }$.
For the type of solutions discussed in section 4, for which we have
a horizon, but for which the exterior region is not Reissner-Nordstrom,
these inequalities can be useful.
There are bounds we can put on the masses and horizon
sizes of such solutions.
{}From Eqs.~(2.12) and (2.24) we have, since $(1/A)'$ is positive,
$$
r_{H}^2 \geq {4\pi G\over {e^2}} (1-{u^*}^2 )^2 \eqn\eqAppiii
$$
With $2G\M(r_{H})=r_{H}$, the left hand
side of inequality \bothbounds\ leads to
$$
M^2 \geq {4\pi \over {Ge^2}} (1-2u^{*2} )\eqno\eq
$$
and
$$
GM-{\sqrt {G^2 M^2 - {4\pi G\over {e^2}}(1-2u^{*2} )}}\leq r_{H} \leq
GM+{\sqrt{G^2 M^2 -{4\pi G\over{e^2}}(1-2u^{*2})}}\eqno\eq
$$
The right hand side of \bothbounds\ does not constrain $r_{H}$ unless
$M^2 \geq {4\pi \over {Ge^2}}(1+ 2u^{*2} -u^{*4} )$, in which case we
get
$$
r_{H}\leq GM-{\sqrt{G^2 M^2 - {4\pi G\over {e^2}}(1+2u^{*2} -u^{*4} )}}
\eqno\eq
$$
or
$$
r_{H}\geq GM + {\sqrt{ G^2 M^2 -{4\pi G\over {e^2}}(1+2u^{*2} -u^{*4} )}}
\eqno\eq
$$

Once we specify the value of $u$ at the horizon, these inequalities
constrain the values of masses and horizon sizes. For example, for the
critical solutions with $u^*$ given by Eq.~(2.13),
inequality \eqAppiii\ gives
$$
e^2 v^2 r_{H}^2 \leq 1-{1\over 2}\left( 1-{{M^2 Ge^2}\over
{4\pi}}\right) \left(1-{e^2 \over \lambda}\right)
\eqno\eq
$$
for $\lambda >e^2 $. For $M^2 \leq {4\pi \over {Ge^2}}$, this is
a refinement of inequality (4.7) in the text. For $\lambda <e^2 $, we get
$$
M^2 \leq {{4\pi (e^2 +\lambda -2\lambda e^2 v^2 r_{H}^2 )}\over {Ge^2
(e^2 -\lambda)}} \eqno\eq
$$
which requires that
$$
\lambda v^2 r_{H}^2 \leq 1-{1\over 2}\left(1-{\lambda \over
e^2}\right)\eqno\eq$$
This is again a refinement of (4.7).

\vfill\eject
\refout
\figout

\end